\documentclass{article}
\usepackage{dcase2020,amsmath,url,times,booktabs,tabularx,float}
\usepackage{amssymb}
\usepackage[ruled,lined]{algorithm2e}
\usepackage{subcaption}
\usepackage{flushend}
\usepackage{bm}

\usepackage[pdftex]{graphicx}

\title{Deep Autoencoding GMM-based Unsupervised Anomaly Detection in Acoustic Signals and its Hyper-parameter Optimization}

\name{Harsh Purohit, Ryo Tanabe, Takashi Endo, Kaori Suefusa, Yuki Nikaido, Yohei Kawaguchi}
\address{ Research and Development Group, Hitachi, Ltd.\\
1-280, Higashi-koigakubo Kokubunji-shi, Tokyo 185-8601, Japan \\
harsh\_pramodbhai.purohit.yf@hitachi.com}

\begin{document}

\ninept
\maketitle

\begin{sloppy}

\begin{abstract}
Failures or breakdowns in factory machinery can be costly to companies, so there is an increasing demand for automatic machine inspection. Existing approaches to acoustic signal-based unsupervised anomaly detection, such as those using a deep autoencoder (DA) or Gaussian mixture model (GMM), have poor anomaly-detection performance. In this work, we propose a new method based on a deep autoencoding Gaussian mixture model with hyper-parameter optimization (DAGMM-HO). In our method, the DAGMM-HO applies the conventional DAGMM to the audio domain for the first time, with the idea that its total optimization on reduction of dimensions and statistical modelling will improve the anomaly-detection performance. In addition, the DAGMM-HO solves the hyper-parameter sensitivity problem of the conventional DAGMM by performing hyper-parameter optimization based on the gap statistic and the cumulative eigenvalues. Our evaluation of the proposed method with experimental data of the industrial fans showed that it significantly outperforms previous approaches and achieves up to a 20\% improvement based on the standard AUC score.
\end{abstract}

\section{Introduction}

 Anomaly events can decrease the quality of manufactured products and deteriorate the reliability of the industrial processes. To avoid this issue, many anomaly detection techniques are applied for the smart factory maintenance. These techniques are mainly based on sensor data parameters, environment variables, quality metrics of the industrial outcomes, and image-based methods \cite{kaiser2009predictive,carino2016enhanced,janssens2015thermal}. Alternatively, recent acoustic scene classification and event detection technologies \cite{lim2017rare,koizumi2018unsupervised} have shown promise for detecting anomalies from sound signals. Our objective in this study, is to develop an acoustic signal-based unsupervised anomaly detection method.

Existing unsupervised anomaly detection methods can be grouped into three categories. (i) Reconstruction-based methods, which assume that anomalies cannot be projected similarly on a low-dimensional space. For example, principle component analysis (PCA) and algorithms based on deep autoencoders (DA) \cite{tagawa2015structured,marchi2015novel,kawaguchi2017can,oh2018residual,salakhutdinov2009deep} have been widely used an achieved acceptable results. However, the performance of these methods is limited because they are based on just one aspect of the reconstruction error. Moreover, low-dimensional representations sometimes loose the essential information of the input data. (ii) Cluster analysis methods like Gaussian mixture models (GMM) \cite{bishop2006pattern} and $k$-means ~\cite{goldstein2016comparative} which, cluster data samples and find anomalies by means of a predefined distance score. However, the performance of these methods is limited by an over-simplified density estimation model that has insufficient capacity. (iii) One-class classification approaches, in which algorithms learn a discriminatory boundary surrounding the normal instances \cite{aurino2014one}. These discriminatory boundary and cluster-based methods can not be directly applied to very high-dimensional data.

Recently, a deep autoencoding Gaussian mixture model (DAGMM) for unsupervised anomaly detection was proposed ~\cite{Zong2018DeepAG}. The DAGMM model utilizes a deep autoencoder to generate a low-dimensional representation and
reconstruction error for each input samples, which is then further fed into a Gaussian mixture model. The parameters of the DA and GMM networks are optimized simultaneously in an end-to-end fashion. However, when applying this method to acoustic anomaly detection, we found that its performance is significantly dependent on two hyper-parameters: (i) the number of GMM components and (ii) the dimension of the compressed features in the autoencoder.

To address these issues, we propose a deep autoencoding Gaussian mixture model with hyper-parameter optimization (DAGMM-HO) and evaluate it in the context of acoustic anomaly detection. This method can determine the anomaly patterns in the time and frequency domains when applied to spectrograms. We propose the following techniques to be used for constructing the DAGMM-HO model:

\begin{itemize}
    \item Gap statistic~\cite{tibshirani2001estimating}-based clustering to determine the number of GMM components
    \item Cumulative eigenvalues in principle component analysis (PCA) \cite{abdi2010principal} to determine the size of the reduced dimension in DA.
\end{itemize}

We evaluated the proposed DAGMM-HO method to see if it could distinguish between normal and abnormal functioning of industrial fans with different sizes and designs. Standard area under curve (AUC) scores and F1 scores were used to evaluate and compare the proposed approach with the conventional methods.

\section{DAGMM-HO-based acoustic anomaly detection model} 
The proposed acoustic anomaly detection model with the deep autoencoding Gaussian mixture model with hyper-parameter optimization (DAGMM-HO) is composed of two parts, as shown in Fig\ref{overview}. In the first part, hyper-parameter tuning has been performed to determine the correct number of GMM components and the optimal dimension for the autoencoder. The second part is the standard DAGMM model with the hyper-parameters selected from the first part.The DAGMM model consists of a \textit{compression} sub-network for dimension reduction using a DA and an \textit{estimation} sub-network that performs a mixture membership prediction in terms of the log-likelihood for the compressed representation of each data sample. Using these predicted membership values, we can estimate the parameters of the GMM through the simultaneous minimization of the reconstruction loss from the compression network and the sample log-likelihood from the estimation network. Before introducing the hyper-parameter optimization methods, we outline the DAGMM  method and related issues. 

\begin{figure}
\begin{center}
\includegraphics[height=4.5 cm,width=8.2cm,clip]{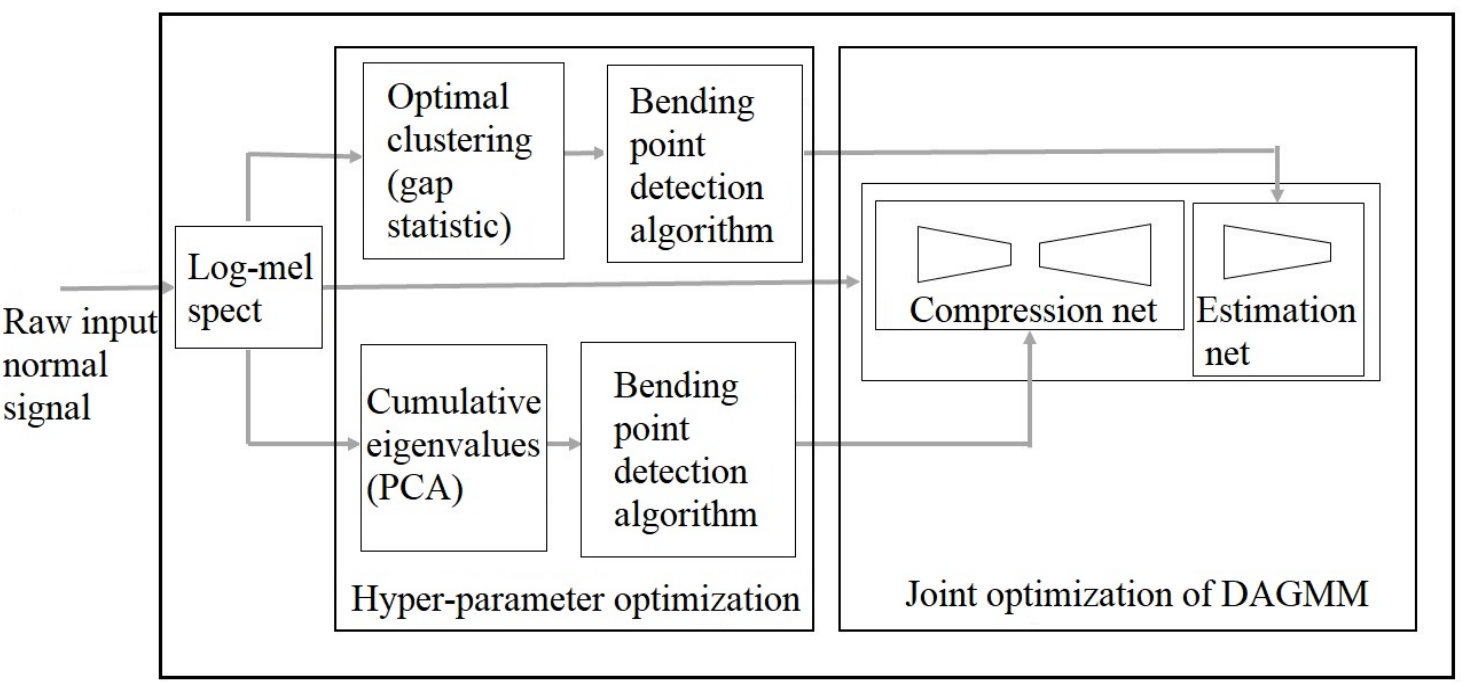}\vspace{-0.3cm}
\caption{Overview of DAGMM-HO-based acoustic anomaly detection.}
\label{overview}
\end{center}
\end{figure}

\subsection{DAGMM model}

The compression network gives a compressed representation $\mathbf{z_c}$ of a given sample $\mathbf{x}$. The feature $\mathbf{z}$ can be defined from two sources: (i) the low-dimensional representation learned by the deep autoencoder and (ii) the feature derived from the reconstruction error. Formally, this is illustrated as 
\begin{align}
    \mathbf{z_c} = f(\mathbf{x}; \theta_{enc}), \\ 
    \mathbf{x'} = g(\mathbf{z_c}; \theta_{dec}),\\
    \mathbf{z_r} = d(\mathbf{x}, \mathbf{x'}),\\
    \mathbf{z}  = [\mathbf{z_c},\mathbf{z_r}],
\end{align}
where $\mathbf{z_r}$ is a potential multi-dimensional feature derived from the reconstruction error, $\theta_{enc}$ and $\theta_{dec}$ are the parameters of the deep autoencoder, $f(\cdot)$ denotes the encoding function, $g(\cdot)$ denotes the decoding function, and $d(\cdot)$ denotes the function for calculating the reconstruction error features, and $\mathbf{x'}$ is the reconstructed counter part of $\mathbf{x}$.

The estimation network performs a density estimation on the output $\mathbf{z}$ from the compression network. We use a multi-layer neural network to predict the mixture membership for each sample, i.e., we estimate GMM parameters in place of using conventional methods like expectation-maximization (EM) \cite{bishop2006pattern}. For the compressed representation $\mathbf{z}$, the estimation network performs the membership prediction as follows:
\begin{align}
    \boldsymbol{\gamma} = softmax(\mathbf{p}),\\
    \mathbf{p} = MLN(\mathbf{z}; \theta_{est}), 
\end{align}
where $\boldsymbol{\gamma}$ is a $K$-dimensional vector for the soft mixture-component membership prediction and $\mathbf{p}$ is the output of a multi-layer network (MLN) parameterized by $\theta_{est}$. Given a batch of $N$ samples and their membership prediction, $\forall 1 \leq k \leq K$, where $K$ is equivalent to the number of GMM components, we can further estimate the parameters in the GMM as
\begin{align}
    \phi_{k} = \sum_{i=1}^{N} \frac{\boldsymbol{\gamma}_{ik}}{N}, \\ \boldsymbol{\mu}_{k} = \frac{\sum_{i=1}^{N}\boldsymbol{\gamma}_{ik}\mathbf{z}_i}{\sum_{i=1}^{N}\boldsymbol{\gamma}_{ik}},  \\
    \mathbf{\Sigma}_{k} = \frac{\sum_{i=1}^{N}\boldsymbol{\gamma}_{ik}(\mathbf{z}_i- \boldsymbol{\mu}_k)(\mathbf{z}_i- \boldsymbol{\mu}_k)^{T}}{\sum_{i=1}^{N}\boldsymbol{\gamma}_{ik}},
\end{align}

where $\boldsymbol{\gamma}_i$ is the membership prediction for the low-dimensional representation $\mathbf{z_i}$, and $\phi_k, \boldsymbol{\mu}_k,$ and  $\boldsymbol{\Sigma}_k$ are the mixture probability, mean, and covariance for component $k$ in the GMM, respectively.
Accordingly, the sample energy/likelihood of an input data sample can be inferred as
\begin{align}
E(\mathbf{z}) = -\text{log}\Big(\sum_{k=1}^{K}\phi_k \frac{\text{exp}\big(-\frac{1}{2}(\mathbf{z - \boldsymbol{\mu}_k})^T\mathbf{\Sigma_k^{-1}}(\mathbf{z} - \boldsymbol{\mu}_k)\big)}{\sqrt{|2\pi\mathbf{\Sigma}_k|}}\Big).
\label{eng}
\end{align}

Given a dataset of $N$ samples, the objective function for joint training of the compression and estimation networks is as follows:
\begin{align}
J(\theta_{enc},\theta_{dec},\theta_{est}) =\frac{1}{N}\sum_{i=1}^{N}L(\mathbf{x}_i,\mathbf{x'}_i) + \frac{\lambda_1}{N}\sum_{i=1}^{N}E(\mathbf{z}_i) \nonumber \\+\lambda_2 P(\mathbf{\Sigma}).
\end{align}
The three main components of the objective function are as follows. (i) $L(\mathbf{x}_i, \mathbf{x'}_i)$ represents the reconstruction error of the deep autoencoder for the given sample $\mathbf{x}$, i.e., $L(\mathbf{x}_i, \mathbf{x'}_i)=\left\lVert \mathbf{x}_i- \mathbf{x}_i'\right\rVert_2^{2}.$
    (ii) The second term shows the energy of the input sample $\mathbf{z_i}$. 
    Minimization of this term leads us to model probabilities from the observed input samples.
    $\lambda_1$ is a hyper-parameter that decides the regularization weight from the estimation network. 
     (iii) The third term is utilised to prevent the singularity problem. Here, $P(\mathbf{\Sigma}) = \sum_{k=1}^K\sum_{j=1}^d \frac{1}{\Sigma_{kj}}$, d is the number of dimensions in the low-dimensional representations provided by the compression network.
Once the model is trained, we can obtained the  output as the estimation of sample energy $E(\mathbf{z}_i)$ and predict the samples of high energy as anomalies in accordance with pre-chosen threshold $\eta$. However, hyper-parameters like $K$ (number of components in GMM) and $c$ (reduced dimension) have to be carefully optimized to ensure high accuracy in the subsequent anomaly detection task.  

\section{ Hyper-parameter optimization}
In order to develop an acoustic anomaly detection method based on DAGMM, we found that two important hyper-parameters should be tuned accurately. Fig \ref{KC} depicts the change in accuracy of anomaly detection with respect to the number of GMM components and reduced dimension in the DA. The black point represents the optimal value for both hyper-parameters at which the anomaly detection system provides the highest accuracy. We propose methods to determine the correct number of clusters in the GMM and optimal dimensions in the DA with the use of normal data as follows. 

\begin{figure}
\begin{center}
\includegraphics[height=6cm,width=8cm]{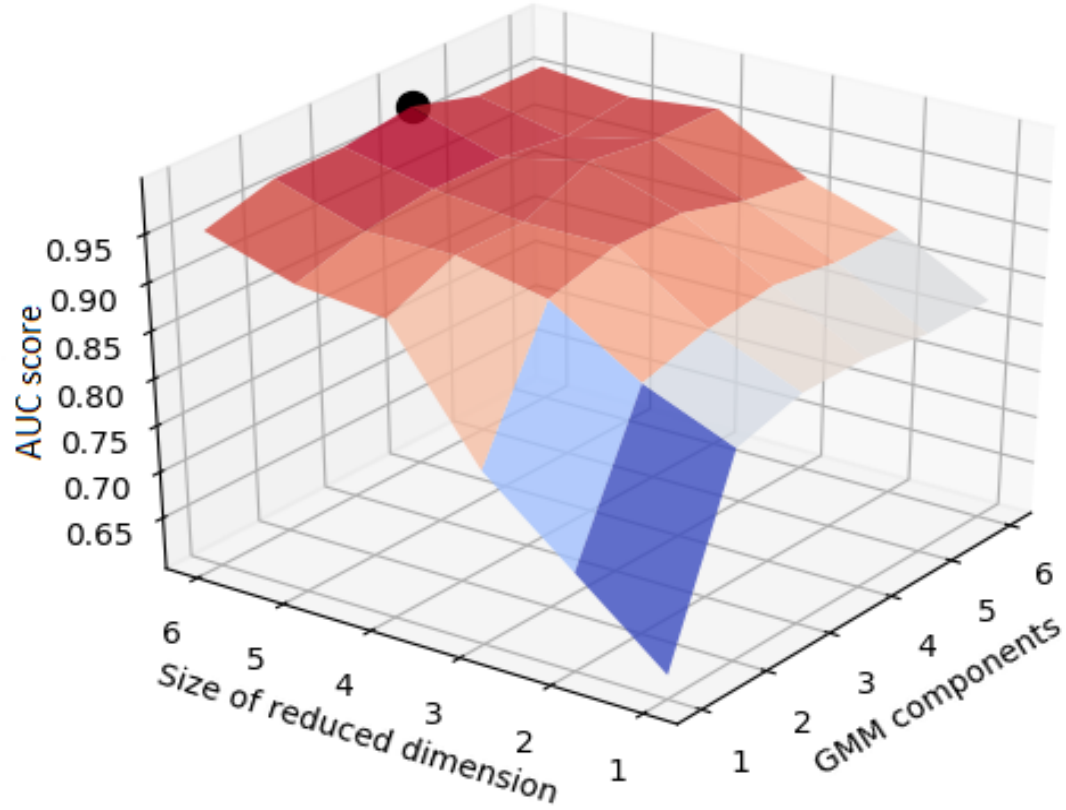}\vspace{-0.3cm}
\caption{Change in accuracy with respect to number of GMM components and  reduced dimensions.} 
\label{KC}
\end{center}
\end{figure}

\subsection{Components in GMM}
\label{sec:gmm_comp}
We propose a method to estimate the optimal number of GMM components using gap statistic \cite{tibshirani2001estimating}. A curve is obtained by calculating the gap statistic for different numbers of GMM components. A bending point representing the maximum change on this curve is used to obtain the optimal component count. The idea behind gap statistic is to find a way to standardize the comparison of variance quantity $\mathcal{D}_o(k)$ with a null reference distribution of the data, i.e., a distribution with no obvious clustering. The gap statistic value $\mathcal{G}_k$ for different cluster counts $k$ is calculated using uniform distribution as a reference with the following equation:

\begin{align}
    \mathcal{G}_k = \mathbb{E}_{n}\big( log(\mathcal{D}_r(k)\big) - log (\mathcal{D}_o(k)),
    \label{eq1}
\end{align}

where $\mathbb{E}_{n}$ denotes expectation under a sample of size $n$ from the reference distribution and $\mathcal{D}_o(k)$ and $\mathcal{D}_r(k)$ represent within-cluster dispersion in the original and reference data, respectively. Within-cluster dispersion $\mathcal{D}(k)$ between points in the given cluster $C_k$ containing points $n_k$ can be calculated as 

\begin{align}
    \mathcal{D}(k) = \frac{1}{2n_k}\sum_{k=1}^{K}\sum_{x_i,x_j \in C_k} \left\lVert x_i - x_j \right\rVert_2^{2}.
    \label{eq2}
\end{align}

 \begin{figure}
    \centering
    \includegraphics[width=0.75\linewidth]{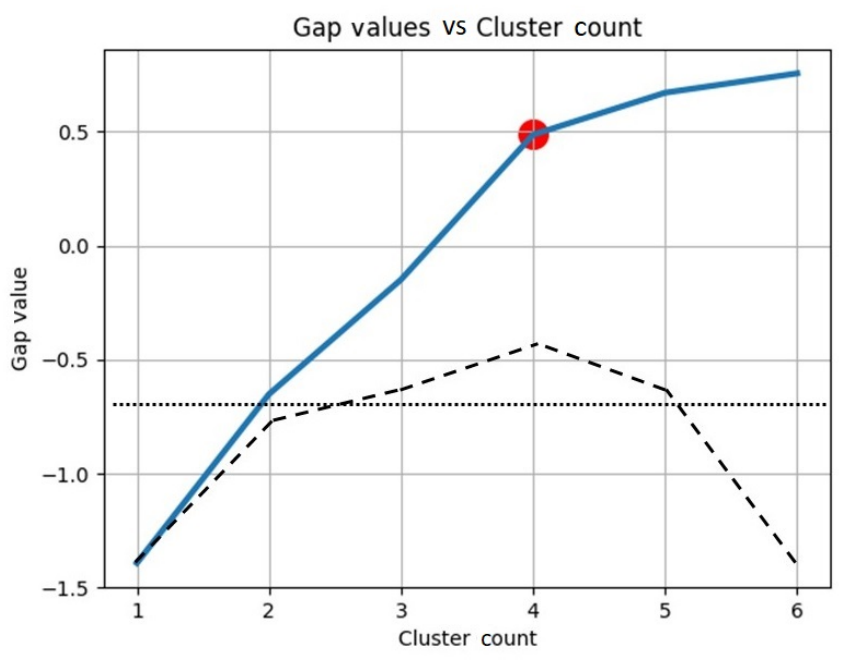}
  \caption{Illustration of curve bending point detection.}
  \label{fig3} 
\end{figure}

The curve in Fig \ref{fig3} represents gap statistic values $\mathcal{G}_k$ for different cluster counts $k$. We are interested in the point of maximum change (represented as the red dot) on the curve because it indicates the best cluster count. For example, in Fig \ref{fan}, six different clusters are present but the best minimum number of components to represent the overall data is four. In order to find this, we need to find the bending point that gives the maximum change in the gap value with respect to the number of clusters. To find this point, we apply Algorithm~\ref{alg2} to the gap curve.

\begin{algorithm}[h]
	\KwIn{Gap statistic values with respective number of clusters $\mathcal{G}_k$}
	\KwOut{Curve bending point $x^{*}$}
\Begin{
		
		 $(x,y)=(k,\mathcal{G}_k)$\\
		 $x_{n_i} = \frac{x_i - min\{x_i\}}{max\{x_i\} - min \{x_i\}} ; \quad y_{n_i} = \frac{y_i - min\{y_i\}}{max\{y_i\} - min \{y_i\}}$\\
		 	
		 $x_{d_i} = x_{n_i} ; \quad y_{d_i} = y_{n_i} - x_{n_i}$\\
		 	
		 $x_{lmx} = x_{d_i}; \quad y_{lmx} = y_{d_i}$ such that $y_{d_{i-1}} < y_{d_i} , ~ y_{d_{i+1}} < y_{d_i}$\\
		 
		 $T_{lmx} = y_{lmx} - \frac{\sum_{i=1}^{m-1}(x_{n_{i+1}} - x_{n_i})}{m-1}$\\
		 If any difference value $(x_{d_j},y_{d_j})$, where $j>i$, drops below the threshold, declare $x_{lmx}=x^{*}$ 
	}
	\caption{Curve bending point detection\label{alg2}}
\end{algorithm}
 
The input of Algorithm \ref{alg2} consists of gap statistic values and the respective number of cluster counts. After fitting the curve to these values, we find the point that gives the maximum change in the gap value with respect to the particular cluster count. In the first step, a smoothing spline is fitted to preserve the shape of the curve ($x,y$) obtained using $\mathcal{G}_k$. Next, the difference values ($x_{d_i}$,$y_{d_i}$) are calculated over the normalized ($x_{n_i}$,$y_{n_i}$) values of the curve (dashed line curve in Fig \ref{fig3}). Local maxima ($x_{lmx}$,$y_{lmx}$) are obtained for this difference curve. A threshold $T_{lmx}$ is defined using these local maxima values (doted line in Fig \ref{fig3}). Local maxima corresponding to the set of difference values $(x_{d_j},y_{d_j})$ greater than this threshold is considered as the maximum change point. The value of $x_{lmx} = x^{*}$ corresponding to this maximum point is the crude yet best estimate of the number of distinct clusters that incorporate similar acoustic characteristics.

\begin{figure}
\begin{center}
\includegraphics[height=6cm,width=8cm]{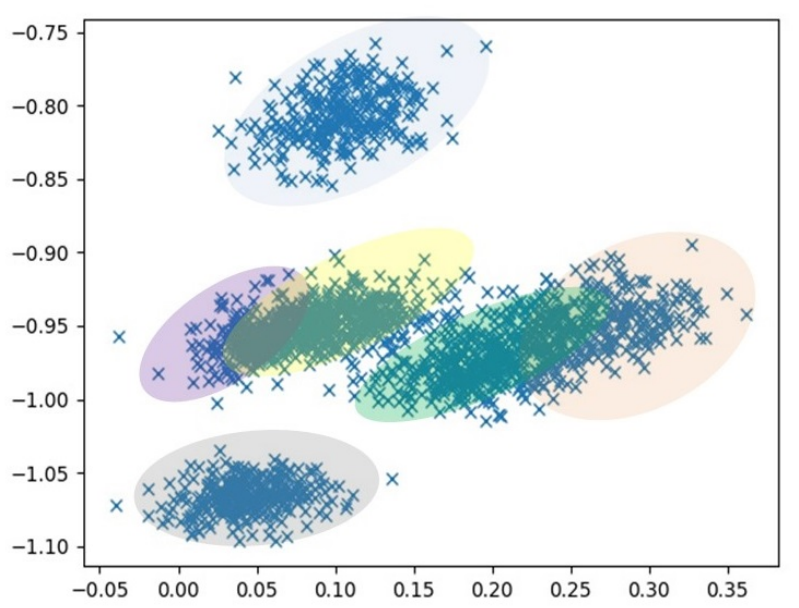}\vspace{-0.3cm}
\caption{2-D representation of audio signals of six industrial fans.} 
\label{fan}
\end{center}
\end{figure}

\subsection{Reduced dimension in DA}
To determine the optimal dimension size at the output of the encoder in the DA, we find the bending point on the curve obtained by taking the cumulative sum of the variance ratio for every component obtained by PCA. We can determine the variance contribution of every component and compress the data w.r.t the number of components required to maintain maximum variance in data using PCA. As described in Section~\ref{sec:gmm_comp}, we use the bending point on the cumulative sum of the variance ratio curve to obtain best estimate of reduced dimension in the DA. 

We obtain a variance ratio $\rho$ by dividing the variance value of each  component $v_i$ by the total variance $v_t$, i.e., the sum of variances of all the components obtained by PCA. Then, a variance ratio curve $\mathcal{V}$ is obtained by taking the cumulative sum of this $\rho$ of every component arranged in descending order. Finally, as described in Algorithm~\ref{alg2}, we find the bending point on the variance ratio curve and determine the optimal size of the encoder output $c$. 

We utilize PCA to estimate the dimension because, in terms of construction, PCA is similar to encoder in that it reduces dimension through the projection of data on principal components. For both, the DA and PCA, the parameters are estimated such that the reconstruction error is minimized, so, PCA can give a good estimate of the reduced dimension size to be used in a neural network-based autoencoder. The DA is flexible in design and non-linearities can be introduced in the model by using different activation functions. Hence, with an increasing amount of features, PCA might be unable to achieve as good a compression as the DA can, but it can be used to obtain an estimate of the reduced dimension.

\section{Experimentation}
To evaluate the effectiveness of the proposed approach, we applied it using real experimental data. We collected sound data ($\sim$ 10 min each) of six industrial fans with different sizes and manufacturers. A microphone was placed at 50 cm away from each fan to collect the audio signal. Saturation of the recorded signal was avoided during data collection. We observed that similar fans generated homogeneous acoustic profiles.

For feature extraction, a log-melspectrogram was calculated from raw audio signals. To compute the mel spectrogram, we considered a frame size of 1024, a hop
size of 512, and 64 mel filter banks.  We predict that for these input features, our trained model will have a low reconstruction error. The DAGMM-HO network structure for the experiment is summarized as follows.

After determining the optimal number of components in the GMM, i.e., $k$, and optimal number of dimensions, $c$, the compression and estimation networks were defined as follows. 
Compression network: $FC(60,30,tanh)$ - $FC(30,10,tanh)$ - $FC(10,c,none)$
- $FC(c,10,tanh)$ - $FC(10,30,tanh)$ - $FC(30,60,tanh)$.
Estimation network: $FC(c+1,10,tanh)$ - $Drop(0.5)$ - $FC(10,k,\textit{softmax})$, 
where $FC(a,b,f)$ means a fully connected layer with $a$ input neurons and $b$ output neurons activated by a function $f$. $Drop(p)$ denotes a dropout layer with a keep probability $p$ during training. Euclidean distance was used to compute the reconstruction error for training.

In the experiments, all the anomalous segments and the same number of normal segments were used as the test dataset and the rest of the normal segments were utilized as the training dataset in our experiment. Energy scores (Equation \ref{eng}) for test-data samples were calculated while predicting from the trained model. These scores were then used to determine the state/health of the machine using a pre-specified threshold.

We compared the proposed approach with the following unsupervised techniques. (i) Deep autoencoder (DA). The deep autoencoder is trained on the normal data and the sample reconstruction error is used as the criterion for anomaly detection. (ii) One-class support vector machine (OC-SVM). OC-SVM is a popular kernel-based method of anomaly detection. In the experiment, we used the radial basis function (RBF) kernel. Note that we provided OC-SVM with an unfair advantage by optimizing its hyper-parameters $\nu$ and $\gamma$. (iii) Gaussian mixture model (GMM) with \textit{diagonal} covariance matrix method. Variants of these methods can be used in a two-step approach. For instance, in step one, we can perform the dimensionality reduction by PCA or by learning the deep autoencoder. Then in step two, we can use OC-SVM or GMM for the classification and density estimation, respectively.

\begin{table}
\centering
\caption{Precision, recall, F1 score and AUC of compared methods\label{CompT}}
\begin{tabular}{ c c c c c}
\hline
\textbf{Method}  &\textbf{Precision} &\textbf{Recall}  &\textbf{F1 score}  &\textbf{AUC}\\

\hline
DAE & $0.68$ &$0.64$& $0.65$&$0.66$\\
GMM& $0.78$ &$0.69$&  $0.75$&$0.77$\\
PCA+GMM& $0.65$ &$0.66$&  $0.65$&$0.69$\\
DAE+GMM& $0.88$ &$0.69$& $0.79$&$0.79$\\
OC-SVM& $0.62$ &$0.85$&  $0.66$&$0.65$\\
PCA+OC-SVM& $0.61$ &$0.83$& $0.65$&$0.63$\\
DAE+OC-SVM& $0.64$ &$0.86$&  $0.67$&$0.67$\\
Proposed& $\mathbf{0.94}$ &$\mathbf{0.93}$&  $\mathbf{0.94}$&$\mathbf{0.96}$\\
\hline
\end{tabular}
\end{table}
The area under the curve (AUC), precision, recall, and F1 score were calculated to evaluate the anomaly detection performance. As shown in Table~\ref{CompT}, the proposed method had a significant improvement in terms of F1 score. Moreover, the two-step method with DA and density estimation performed better compared to the single-step methods and provided up to a 20 $\%$ improvement in AUC.

\section{Conclusions}

   We developed an acoustic anomaly detection method using a deep autoencoding Gaussian mixture model with hyper-parameter optimization. We proposed an automated procedure to determine the optimal number of GMM components and reduce the dimension in the DA. Significant improvement in accuracy over conventional methods was achieved during an experiment with real-time data of industrial fans. The proposed method achieved an AUC score of $96\%$ and F1 score of $94\%$, while the best conventional technique had AUC and F1 scores of only $79\%$. In the future, domain adaptation methods can be used to extend our solution to a wider variety of sensors.

\bibliographystyle{IEEEtran}
\bibliography{refs}

\begin{thebibliography}{10}
\providecommand{\url}[1]{#1}
\def\UrlFont{\rmfamily}
\providecommand{\newblock}{\relax}
\providecommand{\bibinfo}[2]{#2}
\providecommand\BIBentrySTDinterwordspacing{\spaceskip=0pt\relax}
\providecommand\BIBentryALTinterwordstretchfactor{4}
\providecommand\BIBentryALTinterwordspacing{\spaceskip=\fontdimen2\font plus
\BIBentryALTinterwordstretchfactor\fontdimen3\font minus
  \fontdimen4\font\relax}
\providecommand\BIBforeignlanguage[2]{{%
\expandafter\ifx\csname l@#1\endcsname\relax
\typeout{** WARNING: IEEEtran.bst: No hyphenation pattern has been}%
\typeout{** loaded for the language `#1'. Using the pattern for}%
\typeout{** the default language instead.}%
\else
\language=\csname l@#1\endcsname
\fi
#2}}

\bibitem{kaiser2009predictive}
K.~A. Kaiser and N.~Z. Gebraeel, ``Predictive maintenance management using
  sensor-based degradation models,'' \emph{IEEE Transactions on Systems, Man,
  and Cybernetics-Part A: Systems and Humans}, vol.~39, no.~4, pp. 840--849,
  2009.

\bibitem{carino2016enhanced}
J.~A. Carino, M.~Delgado-Prieto, D.~Zurita, M.~Millan, J.~A.~O. Redondo, and
  R.~Romero-Troncoso, ``Enhanced industrial machinery condition monitoring
  methodology based on novelty detection and multi-modal analysis,'' \emph{IEEE
  {A}ccess}, vol.~4, pp. 7594--7604, 2016.

\bibitem{janssens2015thermal}
O.~Janssens, R.~Schulz, V.~Slavkovikj, K.~Stockman, M.~Loccufier, R.~Van~de
  Walle, and S.~Van~Hoecke, ``Thermal image based fault diagnosis for rotating
  machinery,'' \emph{Infrared Physics \& Technology}, vol.~73, pp. 78--87,
  2015.

\bibitem{lim2017rare}
H.~Lim, J.~Park, and Y.~Han, ``Rare sound event detection using 1d
  convolutional recurrent neural networks,'' in \emph{Proceedings of the
  Detection and Classification of Acoustic Scenes and Events (DCASE) Workshop},
  2017, pp. 80--84.

\bibitem{koizumi2018unsupervised}
Y.~Koizumi, S.~Saito, H.~Uematsu, Y.~Kawachi, and N.~Harada, ``Unsupervised
  detection of anomalous sound based on deep learning and the
  {N}eyman--{P}earson lemma,'' \emph{IEEE/ACM Transactions on Audio, Speech,
  and Language Processing}, vol.~27, no.~1, pp. 212--224, 2018.

\bibitem{tagawa2015structured}
T.~Tagawa, Y.~Tadokoro, and T.~Yairi, ``Structured denoising autoencoder for
  fault detection and analysis,'' in \emph{Proceedings of the {A}sian
  {C}onference on {M}achine {L}earning (ACML)}, 2015, pp. 96--111.

\bibitem{marchi2015novel}
E.~Marchi, F.~Vesperini, F.~Eyben, S.~Squartini, and B.~Schuller, ``A novel
  approach for automatic acoustic novelty detection using a denoising
  autoencoder with bidirectional {LSTM} neural networks,'' in \emph{Proceedings
  of the IEEE International Conference on Acoustics, Speech and Signal
  Processing (ICASSP)}, 2015, pp. 1996--2000.

\bibitem{kawaguchi2017can}
Y.~Kawaguchi and T.~Endo, ``How can we detect anomalies from subsampled audio
  signals?'' in \emph{Proceedings of the IEEE 27th International Workshop on
  Machine Learning for Signal Processing (MLSP)}, 2017, pp. 1--6.

\bibitem{oh2018residual}
D.~Oh and I.~Yun, ``Residual error based anomaly detection using auto-encoder
  in smd machine sound,'' \emph{Sensors}, vol.~18, no.~5, p. 1308, 2018.

\bibitem{salakhutdinov2009deep}
R.~Salakhutdinov and G.~Hinton, ``Deep boltzmann machines,'' in
  \emph{Artificial intelligence and statistics}, 2009, pp. 448--455.

\bibitem{bishop2006pattern}
C.~M. Bishop, \emph{Pattern recognition and machine learning}.\hskip 1em plus
  0.5em minus 0.4em\relax {S}pringer, 2006.

\bibitem{goldstein2016comparative}
M.~Goldstein and S.~Uchida, ``A comparative evaluation of unsupervised anomaly
  detection algorithms for multivariate data,'' \emph{PloS one}, vol.~11,
  no.~4, p. 173, 2016.

\bibitem{aurino2014one}
F.~Aurino, M.~Folla, F.~Gargiulo, V.~Moscato, A.~Picariello, and C.~Sansone,
  ``One-class svm based approach for detecting anomalous audio events,'' in
  \emph{2014 International Conference on Intelligent Networking and
  Collaborative Systems}.\hskip 1em plus 0.5em minus 0.4em\relax IEEE, 2014,
  pp. 145--151.

\bibitem{Zong2018DeepAG}
B.~Zong, Q.~Song, M.~R. Min, W.~Cheng, C.~Lumezanu, D.~ki~Cho, and H.~Chen,
  ``{D}eep {A}utoencoding {G}aussian {M}ixture {M}odel for {U}nsupervised
  {A}nomaly {D}etection,'' in \emph{International Conference on Learning
  Representations (ICLR)}, 2018.

\bibitem{tibshirani2001estimating}
R.~Tibshirani, G.~Walther, and T.~Hastie, ``Estimating the number of clusters
  in a data set via the gap statistic,'' \emph{Journal of the Royal Statistical
  Society: Series B (Statistical Methodology)}, vol.~63, no.~2, pp. 411--423,
  2001.

\bibitem{abdi2010principal}
H.~Abdi and L.~J. Williams, ``Principal component analysis,'' \emph{Wiley
  interdisciplinary reviews: computational statistics}, vol.~2, no.~4, pp.
  433--459, 2010.

\end{thebibliography}

%
%
%
%
%
%
%
%
%

\end{sloppy}
\end{document}